\documentclass{phb-proc4-auth}

\usepackage{graphicx}
\usepackage{amssymb}
\usepackage{amsmath}

\begin{document}
\begin{frontmatter}

%----------------------------------------------------------------------
% Specify destination and version number of the manuscript

\journal{SCES '04}

%----------------------------------------------------------------------
% Title of manuscript

\title{Optical conductivity and the sum rule in the DDW state}
%----------------------------------------------------------------------
% List of authors
%
% List each author using a separate \author{} command
%
% If there is more than one author address, add a label to each author
% of the form \author[label]{name}.  This label should be identical to
% the corresponding label provided with the \address command.
%
% e.g. if there are three authors from two institutions in USA and 
% France, you can link them to their respective addresses, using
%
% \author[US]{John Doe}
% \author[US,FR]{Jane Doe}
% \author[FR]{Jean Dupont}
% \address[US]{University of Life, Somewhere, USA}
% \address[FR]{Universite de la Vie, Quelque Part, France}
%
% N.B. Unlike the document class used for abstract submissions, it is
% possible to have the author associated with more than one address,
% as shown in the example above.
%

\author{D. N. Aristov \corauthref{1}\thanksref{ABC} }
\author{R. Zeyher}

%----------------------------------------------------------------------
% List of addresses
%
% If there is more than one address, list each using a separate 
% \address command using a label to link it to the respective author
% as described above
 
\address{
Max-Planck-Institut f\"ur Festk\"orperforschung, Heisenbergstra\ss e 1, 
70569 Stuttgart,
Germany}

%----------------------------------------------------------------------
% Title page footnotes
%
% If you need to add qualifying information to any of the authors, 
% use the \thanksref{} command within the \author command.  The 
% argument is the label of a corresponding \thanks[label]{text}
% command which contains the footnote text
%
% e.g. you can acknowledge a funding authority for John Doe, using
%
% \author{John Doe\thanksref{ABC}}
% \thanks[ABC]{This work was supported by Institute of Unphysical 
%    Phenomena under contract no. ABC-123}
%

%\thanks[]{}

\thanks[ABC]{On leave from 
Petersburg Nuclear Physics Institute, Gatchina  188300, Russia.}

%----------------------------------------------------------------------
% Contact Information
%
% Add the complete postal address, telephone number, fax number, and
% email address of the corresponding author as a special footnote using
% the \corauth[]{} command.  This works in a similar way to the \thanks 
% command.  Add the \corauthref{} command within the \author command.
% The argument is the label of a corresponding \corauth[label]{text}
% command which contains the contact information.  Prefix the text with
% Corresponding Author:
%
% e.g. if the contact author is John Doe,
%
% \author{John Doe\corauthref{1}}
% \corauth[1]{Corresponding Author: University of Life, 123 Some St.,
%    Somewhere, MI 12345, USA.  Phone: (555) 555-5555 
%    Fax: (555) 555-7777, Email: JDoe@uol.edu}
%

\corauth[1]{Corresponding Author : Max-Planck-Institut f\"ur 
Festk\"orperforschung, Heisenbergstra\ss e 1, 70569 Stuttgart,
Germany.   Email: aristov@fkf.mpg.de
}

%----------------------------------------------------------------------
% Text of abstract

\begin{abstract}
The density-wave with $d-$wave order parameter (DDW) is possibly
realized in the underdoped regime of high-$T_c$ cuprates.  The DDW
state is characterized by two branches of low-lying electronic
excitations, and the quantum mechanical current has in particular an
inter-branch contribution. The latter component causes a
finite-frequency response in the optical conductivity and a reduction of 
the Drude contribution. We show that this redistribution of the spectral 
weight leaves the optical sum mostly intact, so that the
restricted optical sum rule is only weakly violated. 
\end{abstract}

%----------------------------------------------------------------------
% Manuscript keywords
%
% Please give two or three keywords in the form: keyword \sep keyword
% e.g. NMR \sep superconductivity
%
% NB The syntax is different from the abstract document class


%----------------------------------------------------------------------
% End of front page

\end{frontmatter}

%----------------------------------------------------------------------
% Manuscript text
\def\bk{{\bf k}}
\def\bQ{{\bf Q}}
\def\ve{\varepsilon}

Recently, the interest in charge density waves with unconventional
order parameters has increased \cite{Maki-Neto,Cappelluti,Chakravarty}.
In particular, it has been shown that a charge density wave with d-wave
symmetry (DDW) represents a stable state of the $t-J$ model in the
large-N limit in certain doping and temperature regions\cite{Cappelluti}.
It thus may be intimately related to the pseudogap phase of 
high-T$_c$ superconductors\cite{Cappelluti,Chakravarty}. 
The presence of a DDW state should also cause
changes in transport coefficients\cite{WKim02}.
In the previous paper \cite{ArZe-hall} we showed that the usual expression for 
the Hall conductivity is modified in the DDW state, due to specific matrix 
element for the current operator.
In the present paper we show that this interband current is responsible for the 
optical response at higher energies and the reduction of the Drude contribution.

The mean-field Hamiltonian in the DDW state is 
       \begin{eqnarray}
       { H}&=& \sum_{{\bf k},\sigma}
       \left[
       \xi_{\bf k} a^\dagger_{{\bf k}\sigma} a_{{\bf k}\sigma}
       + i\Delta_{\bf k} a^\dagger_{{\bf k}\sigma} a_{{\bf k}+{\bf Q},\sigma}
        +h.c. \right]
       \label{H}
       \end{eqnarray}
Considering nearest and next-nearest neighbor hoppings $t$ and $t'$
and putting the lattice constant of the square lattice
to unity, the electronic
dispersion is $\xi_{\bf k}= -2t (\cos k_x +\cos k_y) +4t' \cos k_x
 \cos k_y -\mu$ ; we use the abbreviations
$ \xi_\pm = (\xi_{\bf k} \pm \xi_{{\bf k}+{\bf Q}})/2$ below.
The DDW order parameter is of
the form $\Delta_{\bf k} = \Delta_0(\cos k_x - \cos k_y) = 
- \Delta_{{\bf k}+{\bf Q}}$ ; here ${\bf Q} ={(\pi,\pi)}$.

The Hamiltonian (\ref{H}) can be diagonalized \cite{ArZe-hall} by a
unitary transformation $U$ so that  $\hat h \equiv U \hat H U^\dagger $ is 
diagonal and the new quasiparticle energies are
       \begin{equation}
       \varepsilon_{1,2}= \xi_{+} \pm
       \left[ \xi_{-}^2 +\Delta_{\bf k}^2 \right]^{1/2}.
       \end{equation}

The current operator in the DDW state is defined by differentiating $\partial H 
/\partial k_\alpha =\partial(U^\dagger \hat h U) /\partial k_\alpha$. In the new 
quasiparticle basis it has the components             
            \begin{eqnarray}
             v_{1(2)}^\alpha &=& \frac{\partial
           \varepsilon_{1(2)}}{\partial  k_\alpha}, \quad
           v_{3}^\alpha =
	   \frac{\xi_{-}^2}{({\xi_{-}^2+\Delta_{\bf k}^2})^{1/2}}
	    \frac{\partial}{\partial k_\alpha} \frac{\Delta_{\bf k}}{\xi_{-}}.
            \label{velocities}
	    \end{eqnarray}
Here the velocities $v_{1(2)}$ correspond to the intraband current operator in 
the corresponding subband, $\varepsilon_{1(2)}$. 
The off-diagonal current $v_3$ arises from the $\bf k$-dependence of the 
unitary transformation $U$, and is the matrix element of the total current 
operator connecting the two subbands $\varepsilon_{1}$  and $\varepsilon_{2}$. 
It corresponds to the interband transition operator  $\Omega$ in the notation of 
\cite{LL9} and is needed for a proper description of the optical response.
    
The frequency-dependent conductivity has the form 
$\sigma_{\alpha}(\omega) =\sigma_{\alpha}^{(D)}(\omega) + 
\sigma_{\alpha}^{(opt)}(\omega)$, where  

      \begin{equation}
      \sigma_{\alpha}^{(D)} =
      \frac{e^2 \tau}{1+ \omega^2 \tau^2}  \int \frac{d^3\bk} {(2\pi)^3} 
     \left [
       ({ v}_{1}^\alpha)^2
        (-\frac{\partial n(\ve_{1})}{\partial \ve_{1}}) + (1\leftrightarrow2)
      \right],
       \nonumber
      \end{equation}
is the Drude part of $\sigma_\alpha(\omega)$, written for a model with    
point-like impurities,   and           
      \begin{eqnarray}
      \sigma_{\alpha}^{(opt)} &=&
      \frac{\pi e^2 }{\omega}  \int \frac{d^3\bk} {(2\pi)^3} 
       [n(\ve_{1}) - n(\ve_{2})] 
       ({ v}_{3}^\alpha)^2
       \delta(\omega + 
       \varepsilon_{1}-\varepsilon_{2})   
       \nonumber
       \\ &&
        + (1\leftrightarrow2)
      \end{eqnarray}
is the optical conductivity. Here $\tau$ stands for the (large)
scattering time, $ n(x) = (e^{x/T}+1)^{-1}$. 
      
\begin{figure}[tp]
\begin{center}
\includegraphics[width=0.5\textwidth]{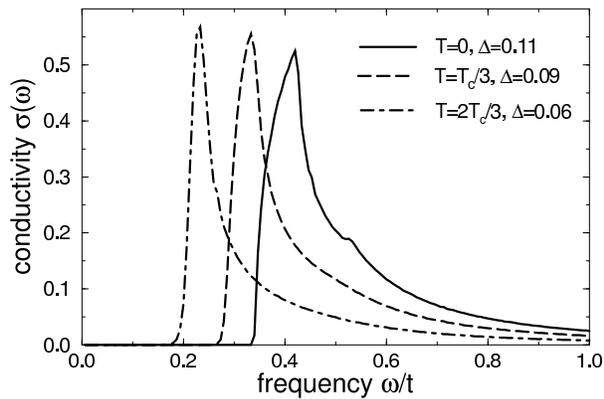}\\
\end{center}
\caption{\label{fig:opt} The evolution of the shape of the optical response 
below the DDW transition. The $\delta-$function-like Drude peak is not shown 
here.}
\end{figure}
%\vspace*{-4mm}

In rough agreement with \cite{Cappelluti}, we modelled the gap by 
$\Delta_0 = \bar{\Delta}(T)\sqrt(1+\mu)\Theta(1+\mu)$, where $\mu$ is the 
chemical potential, $\bar{\Delta}(0) = 0.29 t$, a BCS temperature
dependence is assumed for $\bar{\Delta}(T)$, and $t$ is used as the energy 
unit. 
We show the results for the optical conductivity $\sigma_{x}^{(opt)}$ in 
Fig.\ (\ref{fig:opt}) for $t'=0.3$, $\mu =-0.85$ (doping $\delta\simeq0.08$), 
$T_c =0.064$ and various temperatures.  

First,  the optical response is non-zero only above the threshold energy 
$2\Delta_{hs}$, with  $\Delta_{hs}$ being the value of the gap at the so-called 
``hot-spots'' in $\bk-$space where $\xi_{\bk}\simeq \xi_{\bk + \bQ}\simeq 0$.  
One has roughly $2\Delta_{hs} \sim 4\Delta_0$ for the above parameters. Beyond 
its maximum at $\omega \sim 4\Delta_0$, one observes the decay  
$ \sigma_{x}^{(opt)} \propto \omega^{-3}$ up to energies of the order of the
bandwidth.  The weight associated with $ \sigma_{x}^{(opt)}$ is $ \sim 
|\Delta_0|/t $ as is suggested by the analysis below.

\begin{figure}[tp]
\begin{center}
\includegraphics[width=0.5\textwidth]{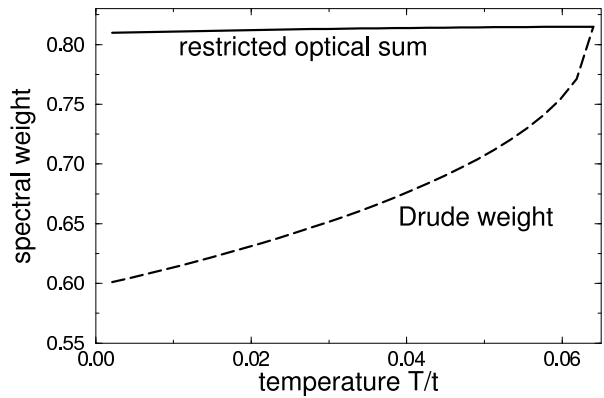}\\
\end{center}
\caption{\label{fig:weight} 
The temperature dependence of the optical sum and the Drude weight below the DDW 
transition at $T_c \sim 0.064 t$ }\end{figure}

The restricted optical sum is given by the integral $\int d\omega 
\sigma_x(\omega)$ which can be easily evaluated for the above 
$\sigma_{\alpha}^{(D)}$, $ \sigma_{\alpha}^{(opt)}$. It was shown in 
\cite{ArZe-hall} that this sum should exhibit a deviation  
$\sim \Delta^2/E_F$ in the DDW state. At the same time the change in the weight 
of the Drude peak, $\pi \tau^{-1} \sigma_{\alpha}(\omega=0)$, is a first-order 
effect in the DDW state, $\delta\sigma_{x}(\omega=0)\sim \Delta/E_F$. 
\cite{ArZe-hall} We illustrate these general findings in Fig.\ 
(\ref{fig:weight}).  It is seen that the restricted optical sum is nearly 
unchanged by the formation of the DDW gap, but that the Drude weight 
exhibits a significant variation below $T_c$.

%\vspace*{-4mm}

Our results for the optical sum differ from those in \cite{OCSR}, because there
a different spectrum with $t'= 0$ and a much larger value of $\Delta_0(T=0) 
\sim t$ were used.

In summary, we have shown that the DDW state leads to a shift of the optical 
spectral weight to higher energies $\sim 4\Delta_0$, but to no change in the 
optical sum rule in leading order. 

%----------------------------------------------------------------------
% Reference section
%
% \bibitem{TdB2}
% J. Doe, J. Doe, and J. Dupont, J. Irrep. Res. 10 (2000) 1000.

%----------------------------------------------------------------------
% Figures and Tables
%
%
% \begin{figure}
%     \centering
%     \includegraphics{filename.eps}
%     \caption{Insert figure caption here} 
% \end{figure}  
%

\end{document}